\documentclass[preprint,12pt]{elsarticle}

\usepackage[T1]{fontenc}
\usepackage[utf8]{inputenc}
\usepackage{amsmath}
\usepackage{bm}
\usepackage{booktabs}
\usepackage{longtable}
\usepackage{tabularx}
\usepackage{array}
\usepackage{graphicx}
\usepackage{algorithm}
\usepackage{algpseudocode}
\usepackage{hyperref}
\usepackage{ragged2e}
\usepackage{xcolor}
\usepackage{diagbox}

\newcolumntype{Y}{>{\RaggedRight\arraybackslash}X}
\biboptions{sort&compress}

\begin{document}

\begin{frontmatter}

\title{A data-driven heuristic-ensemble method for behavioural inference in collective-risk dilemma experiments}

\author[ub,ubics]{Marc Sadurn\'i}
\author[ub,ubics]{Mart\'in F. D\'iaz}
\author[ub,ubics]{Josep Perell\'o\corref{cor}}\ead{josep.perello@ub.edu}\cortext[cor]{Corresponding authors:}
\author[ub,ubics,SH]{Miquel Montero\corref{cor}}\ead{miquel.montero@ub.edu}
\affiliation[ub]{organization={Departament de F\'isica de la Mat\`eria Condensada, Universitat de Barcelona}, addressline={Mart\'i i Franqu\`es, 1}, city={Barcelona}, postcode={E-08028}, state={Catalonia},country={Spain}}
\affiliation[ubics]{organization={Universitat de Barcelona Institute of Complex Systems (UBICS), Universitat de Barcelona},city={Barcelona},state={Catalonia},country={Spain}}
\affiliation[SH]{organization={Serra H\'unter Fellow, Generalitat de Catalunya},state={Catalonia}, country={Spain}}


\begin{abstract}
The collective-risk dilemma has become a paradigm for studying cooperation in a wide variety of social experiments, initiatives that generate increasingly rich datasets describing individual actions over successive decision-making processes. Beyond the purely descriptive analysis of such empirical datasets, a methodological framework is required to interpret and reproduce the observed dynamics, a scheme that may be skewed by exogenous biases. Here, we introduce a data-driven, non-parametric method for simulating individual decisions together with a validation protocol for assessing the fidelity level of the outcomes they generate. Our methodological contribution is based on an empirical heuristic ensemble: agents are represented by a finite set of behavioural heuristics, probabilities conditioned on different observable features of the game state, endowed with a hard-voting arbitration that determines their per-round final decisions. This reproducible generative framework allows for behavioural inference in different collective-risk dilemma setups.
\begin{itemize}
    \item Behavioural heuristics are seeded directly from empirical data assessing conditional probabilities, without a priori assumptions.
    \item A plurality vote with uniform random tie-breaking resolves between competing heuristics each round, adding no free parameters.
    \item Validation follows a pattern-oriented approach, comparing synthetic and empirical macro-indicators against single-heuristic and uniform-random baselines.
    \item Synthetic data generated allow exploration of scenarios or experimental setups that have not been empirically realised.
    \item As a worked example, we instantiate our methodological approach on the \textit{xAire} citizen science collective-risk dilemma experiment run in Barcelona.
\end{itemize}
\end{abstract}


\end{frontmatter}


\section*{Background}

Collective-risk dilemmas (CRDs) have become a central experimental framework for studying cooperation in situations where individuals must contribute private resources toward a collective target while facing the possibility of a shared loss if the target is not achieved \cite{Milinski2008}. Originally developed to investigate climate-change mitigation, CRDs capture the tension between individual incentives to free-ride and the collective need for coordinated action. The seminal experiment by Milinski et al. demonstrated that groups can coordinate contributions when collective failure entails substantial risk \cite{Milinski2008}. Subsequent studies showed that cooperative outcomes depend strongly on contextual factors such as inequality in resource distributions \cite{Tavoni2011,Vicens2018,Milinski2011}, uncertainty regarding collective risks \cite{Barrett2012}, and intergenerational considerations \cite{Jacquet2013}. Together, these experiments generated increasingly rich datasets describing individual contribution dynamics over repeated interactions.

In parallel, CRDs have inspired a substantial body of theoretical and computational research aimed at understanding the mechanisms that produce cooperative outcomes. The risk of collective failure fundamentally alters strategic incentives and can promote cooperation beyond what is expected in standard public-goods settings \cite{SantosPacheco2011}. Subsequent work extended this perspective through evolutionary and agent-based approaches, highlighting the importance of adaptation, social influence, and population heterogeneity in shaping collective outcomes \cite{Santos2012}. Studies incorporating inequality further demonstrated how differences in available resources affect both individual behaviour and collective success \cite{Vasconcelos2014}.

More recently, experimental studies have provided additional evidence that behavioural heterogeneity plays a key role in collective-risk settings. Resource heterogeneity can generate highly unequal effort distributions even when groups achieve collective goals \cite{Vicens2018}. Likewise, long-term collective-risk experiments demonstrated that repeated exposure to collective risks can modify social norms and promote sustained cooperation \cite{Szekely2021}. These findings suggest that aggregate outcomes such as total contributions, target achievement, or payoff distributions provide only a partial picture of collective action, as similar outcomes may emerge from distinct behavioural pathways.

The growing availability of high-resolution experimental data creates new opportunities to investigate these mechanisms directly. CRD experiments routinely record individual decisions across multiple rounds, making it possible to analyse how participants respond to their own previous actions, the behaviour of others, collective progress toward a target, and changes in their available resources. This is the case of the \textit{xAire} citizen science experiment, conducted as a pop-up intervention in two public spaces in Barcelona using portable digital interfaces \cite{Vicens2018Experiment,Sagarra2016}. The experiment generated detailed state--action records from CRD games under different resource distributions and inequality treatments, providing a rich empirical basis for behavioural inference \cite{Vicens2018xairedata}. Experimental and theoretical evidence in different experiments thus suggests that individuals may simultaneously react to several feedback channels, exhibiting persistence in their own behaviour, conditional responses to others, sensitivity to collective progress, and adaptation to resource availability \cite{Tavoni2011,SantosPacheco2011,Vasconcelos2014}. Consequently, observed decisions may result from the interaction of multiple behavioural heuristics rather than from a single dominant strategy.

The methodology presented here \cite{sadurni2026emergent} addresses this challenge by adopting an inference-oriented perspective in which behaviour is characterised through probabilistic relationships between observable states and subsequent actions. Rather than assigning participants to predefined behavioural classes or specifying strategy distributions \emph{a priori}, the method estimates empirical conditional action probabilities directly from experimental observations. All behavioural parameters are inferred from observed state--action records, eliminating the need for external calibration data or pre-specified behavioural models. The underlying premise is that recurring behavioural patterns can be represented as state-conditioned decision rules linking actions to feedback available during the interaction, including individual history, social information, collective progress, and resource availability. In addition to identifying behavioural traits directly from empirical evidence, our methodological approach enables us to generate large sets of synthetic data, allowing for the minute exploration of tested scenarios or the consideration of experimental configurations beyond those that have actually been implemented.

As a worked example, we apply the methodology to the \textit{xAire} CRD dataset \cite{Vicens2018xairedata}. The dataset contains sequences of contribution decisions together with information on players' previous actions, collective progress, and resource availability, making it well suited for identifying behavioural regularities and evaluating alternative behavioural hypotheses (heuristics in the forthcoming sections). Although illustrated here using \textit{xAire} and CRD experiments, the same perspective can be applied to other behavioural experiments involving (unique or successive) decisions and observable contextual variables. By identifying empirical probabilistic links between states and actions, the methodology provides a transparent and reproducible approach to characterising behavioural mechanisms, comparing behavioural heuristics, and building generative models grounded in observations.

\section*{Method details}

The method generates synthetic populations of agents that play a CRD and reproduces a set of emergent macro-indicators of an empirical system. It follows a four step workflow: (i) the identification of a set of behavioural heuristics, $\{H_0,\dots,H_m\}$, each defined as a contribution probability conditioned on one observable feature of the game state; (ii) the estimation of those conditional probability distributions directly from experimental data; (iii) a simulation pipeline in which agents sample contributions from the estimated distributions, either under a single-heuristic or under an ensemble that combines all heuristics through a plurality vote; and (iv) a multi-pattern validation protocol comparing synthetic and empirical macro-indicators. Empirical sampling and hard voting are both standard in the literature; what we add in this method is the way the behavioural heuristics are put together to decide one action from several competing heuristics. 

\begin{figure}[tb]
\centering
\includegraphics[width=\textwidth]{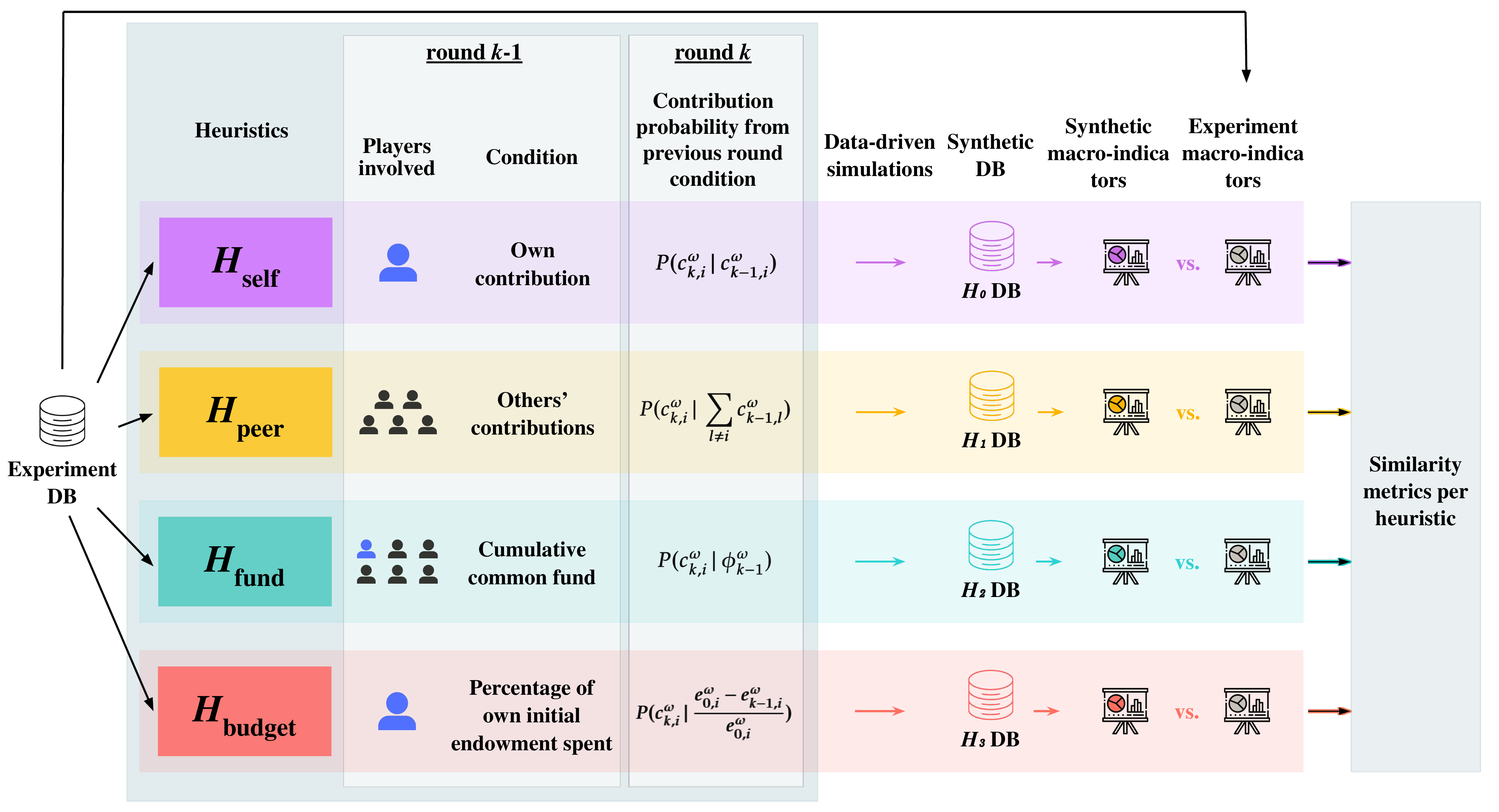}
\caption{Per-heuristic pipeline of the empirical heuristic-ensemble (EHE) method. From the experiment database, each heuristic reads one feature of the round-$(k{-}1)$ state and turns it into a contribution probability for round $k$. Sampling these probabilities produces a synthetic database per heuristic, whose macro-indicators are compared with the experimental ones through similarity metrics. The worked example uses four heuristics ($H_{\mathrm{self}}, H_{\mathrm{peer}}, H_{\mathrm{fund}}, H_{\mathrm{budget}}$). The ensemble combines all four each round through the plurality vote shown in Fig.~\ref{fig:vote}.}
\label{fig:pipeline}
\end{figure}

\begin{figure}[tb]
\centering
\includegraphics[width=\textwidth]{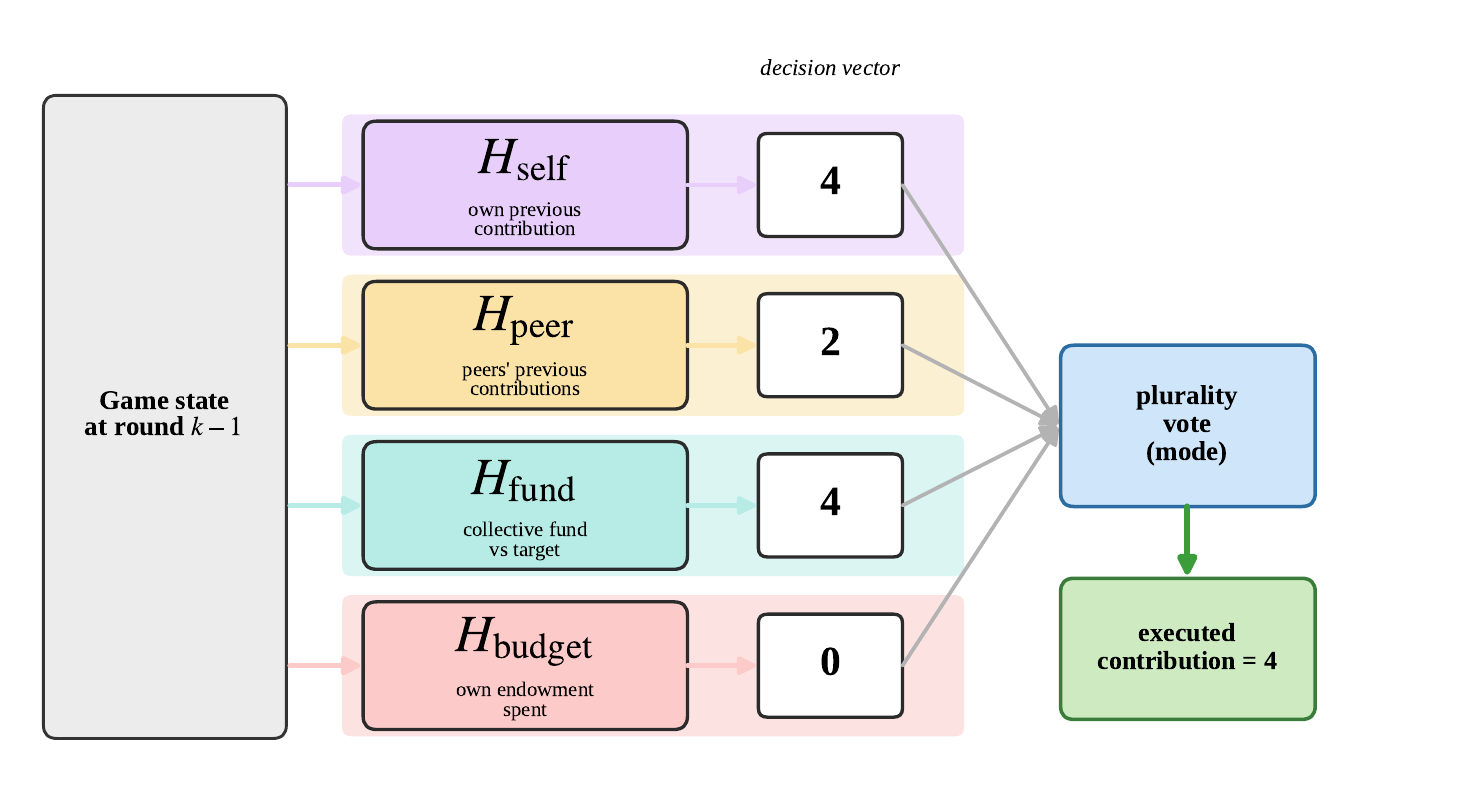}
\caption{Ensemble decision rule for one round. Each heuristic reads the round-$(k{-}1)$ state and proposes a contribution; the four proposals form the decision vector, and the executed contribution is its plurality (mode). When two values tie, one is drawn uniformly at random.}
\label{fig:vote}
\end{figure}

The workflow is summarised in Fig.~\ref{fig:pipeline}, which displays the per-heuristic route, while the dynamics of the ensemble is illustrated in Fig.~\ref{fig:vote}, where it is shown how the plurality arbitrator integrates and resolves the per-round suggestions of the different heuristics. The four heuristics used in these figures are one \emph{instantiation}, chosen because they map onto the feedback channels the CRD interface presented to participants in the \textit{xAire} experiment; a different game with a different observable state would instantiate a different set. The four heuristics chosen here are therefore a feature of the worked example, not a limitation of the method.

It is important to note that this presented framework is \emph{dataset-agnostic}: it takes any threshold-public-goods or collective-risk dataset with per-round, per-participant contribution logs, the initial composition of the game and other publicly available information, and returns a calibrated generative model. All behavioural parameters are estimated from these logs; no external data sources or guidelines are needed.

\subsection*{Game setting and notation}

In the \textit{xAire} experiment used here and hereafter as a worked example, each game involves $N=6$ players interacting over $R=10$ rounds. These values are specific to the illustrative dataset and are not required by the general methodology. In round $k$, player $i$ chooses a contribution $c^{\,\omega}_{k,i}\in\{0,2,4\}$ monetary units (MU) from the remaining endowment, where $\omega$ indexes the wealth treatment to be specified below. The collective target is half the total initial endowment,
\begin{equation}
T=\tfrac{1}{2}\sum_{i=1}^{N} e_{0,i}^{\,\omega}=120~\text{MU},\qquad 
\sum_{i=1}^{N} e^{\,\omega}_{0,i}=240~\text{MU in every treatment.}
\end{equation}
The cumulative common fund and the evolving individual endowment are
\begin{equation}
\phi^{\,\omega}_{k}=\sum_{k'=1}^{k}\sum_{i=1}^{N} c^{\,\omega}_{k',i},
\qquad
e^{\,\omega}_{k,i}=e^{\,\omega}_{0,i}-\sum_{k'=1}^{k} c^{\,\omega}_{k',i}.
\end{equation}
In the \textit{xAire} experiment, three different group compositions were considered; based on this, the initial endowment of each player is assigned under one of the following treatments that hold the initial total endowment fixed at 240~MU as demanded: the Equal treatment ($\omega=\text{E}$, $40$~MU for all), the Unequal-L treatment ($\omega=\text{L}$, two players at $24$~MU, four at $48$~MU) and the Unequal-H treatment ($\omega=\text{H}$, four at $30$~MU, two at $60$~MU). It is important to note that, in our case, the same initial endowment is not present in more than one treatment.

\subsection*{State-conditioned contribution heuristics}\label{sec:heuristics}
Each heuristic is linked to a probability distribution over the action set $\{0,2,4\}$~MU, conditioned on one observable feature of the round-$(k{-}1)$ game state. All four are therefore first-order Markovian on different projections of the state; they differ in \emph{which} state they read, not in their memory order. The worked-example set is:
\begin{description}
  \item[$H_{\mathrm{self}}$ -- own-action heuristic.] Conditioned on the player's own contribution in the previous round, $c^{\,\omega}_{k-1,i}$. As a lookup table this is a $3\times3$ transition matrix over $\{0,2,4\}$~MU.
  \item[$H_{\mathrm{peer}}$ -- peer heuristic.] Conditioned on the mean contribution of the other $N{-}1$ players in the previous round, $\tfrac{1}{N-1}\sum_{j\neq i} c^{\,\omega}_{k-1,j}$ (social information).
  \item[$H_{\mathrm{fund}}$ -- progress heuristic.] Conditioned on the cumulative common fund $\phi^{\,\omega}_{k-1}$ relative to the target $T$ (collective progress).
  \item[$H_{\mathrm{budget}}$ -- budget heuristic.] Conditioned on the fraction of the player's own initial endowment already spent (own resource state).
\end{description}

\subsection*{Estimating the conditional distributions from data}

\begin{figure}[tb]
\centering
\includegraphics[width=\textwidth]{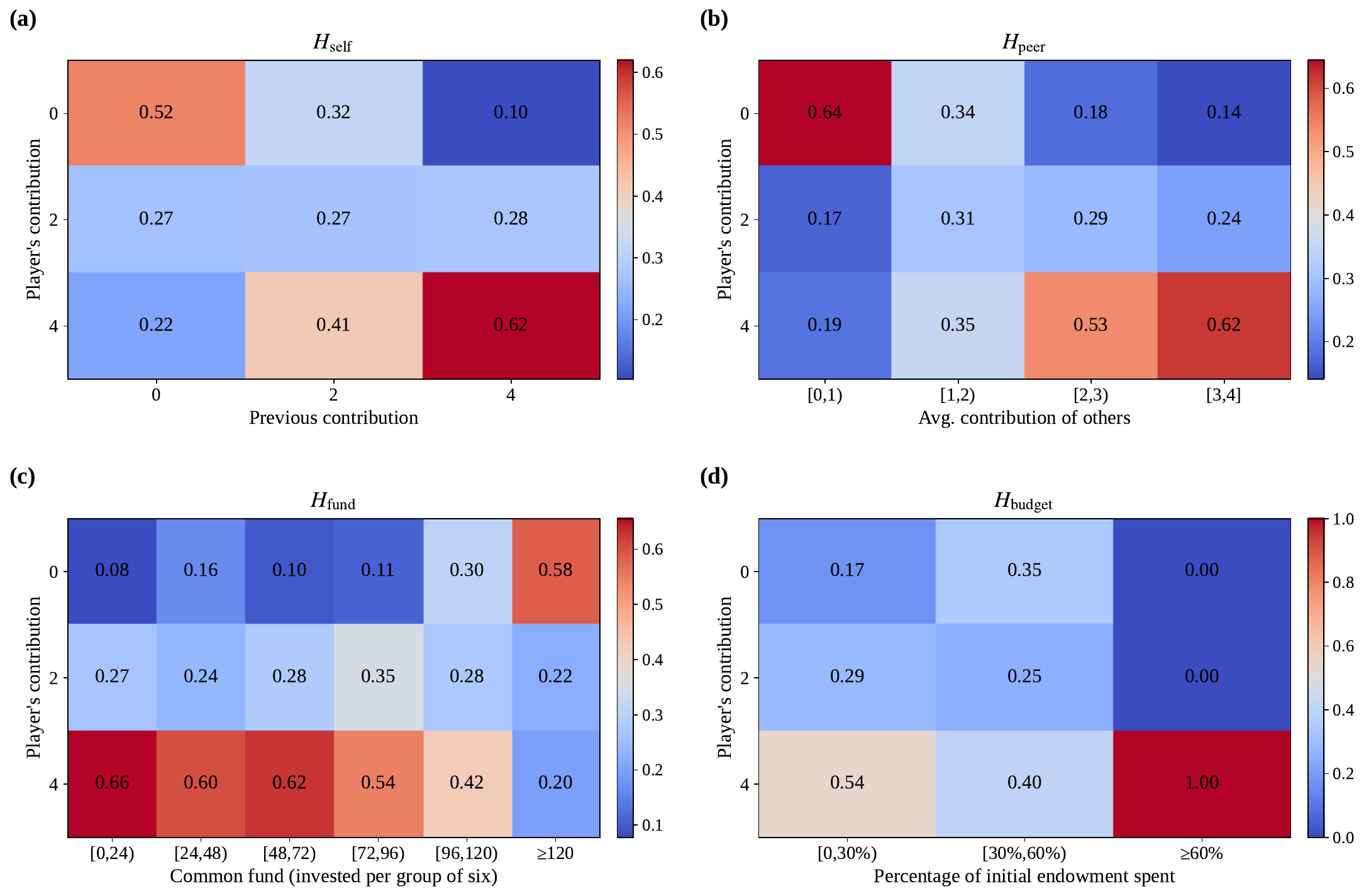}
\caption{Estimated conditional contribution probabilities for the four heuristics, shown at one representative endowment ($60~\mathrm{MU}$). Each cell gives $P(\mathrm{contribution}=0,2,4~\mathrm{MU}\mid\text{conditioning bin})$. (a) $H_{\mathrm{self}}$, conditioned on the player's own contribution in the previous round: a $3\times3$ transition matrix over $\{0,2,4\}~\mathrm{MU}$. (b) $H_{\mathrm{peer}}$, conditioned on the mean contribution of the other five players in the previous round. (c) $H_{\mathrm{fund}}$, conditioned on the cumulative contribution in the common fund relative to the $120~\mathrm{MU}$ target. (d) $H_{\mathrm{budget}}$, conditioned on the fraction of the player's initial endowment already spent. Colour scales are set per panel; the full per-endowment set is reproducible from the released GitHub repository \href{https://github.com/mf-diaz/Project-xAire-EHE}{Project-xAire-EHE}.}
\label{fig:heatmap}
\end{figure}

For each heuristic, the conditioning variable is discretised into bins; the empirical probability of each action $\{0,2,4\}$~MU is computed within each bin, stratified by treatment and initial-endowment level and aggregated across rounds and games. The bins are: previous own contribution, taking the three values $\{0,2,4\}~\mathrm{MU}$, for $H_{\mathrm{self}}$, Fig.~\ref{fig:heatmap}(a); the mean contribution of the other five players in the previous round, binned as $[0,1)$, $[1,2)$, $[2,3)$, and $[3,4]~\mathrm{MU}$, for $H_{\mathrm{peer}}$, Fig.~\ref{fig:heatmap}(b); the cumulative common fund at the end of the previous round, binned in six intervals of $24~\mathrm{MU}$ from $[0,24)$ up to $[96,120)$, with a final open bin $\geq120~\mathrm{MU}$ covering groups that have already met the target, for $H_{\mathrm{fund}}$, Fig.~\ref{fig:heatmap}(c); and the fraction of the player's own initial endowment already spent, binned as $[0,30\%)$, $[30\%,60\%)$, $\geq60\%$, for $H_{\mathrm{budget}}$, Fig.~\ref{fig:heatmap}(d). The result is, per heuristic, a lookup table mapping (endowment, treatment, conditioning bin) to a probability vector over the action set as shown in Fig.~\ref{fig:heatmap}. The first contribution for each synthetic player is sampled directly from the empirical distribution of observed first-round actions, stratified by initial endowment (Table \ref{tab:firstcon}). These empirical frequencies are drawn from both \textit{successful} and \textit{unsuccessful} real games to capture the full spectrum of decision-making tendencies, and to avoid introducing selection bias in the simulation.

\begin{table}[bt!]
    \centering
    \caption[Empirical distribution of first-round contributions.]{\textbf{Empirical distribution of first-round contributions.} Probability of contributing 0, 2, or 4 MU in round $k=1$, based on observed frequencies in the experimental dataset. These values are used to initialise synthetic player behaviour at the start of each game.}
    \begin{tabular}{c|c|c|c}
    \toprule
    \diagbox{\textbf{Endowment (MU)}}{\textbf{Contribution (MU)}} & 0 & 2 & 4 \\ \midrule
    24 & 0.07 & 0.50 & 0.43 \\ \hline
    30 & 0.05 & 0.61 & 0.34 \\ \hline
    40 & 0.06 & 0.50 & 0.44 \\ \hline
    48 & 0.03 & 0.55 & 0.42 \\ \hline
    60 & 0.05 & 0.35 & 0.60 \\ 
    \bottomrule
    \end{tabular}
    \label{tab:firstcon}
\end{table}

\subsection*{Simulation pipeline}
Under a single-heuristic simulation every agent follows one heuristic for all rounds; under the ensemble every agent evaluates all heuristics each round and arbitrates. The procedure is given in Algorithm~\ref{alg:sim}, which is written for an arbitrary number of games and treatments; the two concrete run sizes used in this worked example are stated here first, since both are referred to throughout the rest of the paper. The \emph{matched run} contains the same number of games and treatment configurations as the original dataset, $77$ games split across the three treatments, and is the basis for the scores in Table~\ref{tab:val}. A larger run of $30{\,}000$ games, $10{\,}000$ per treatment, the \emph{extensive run}, is used to characterise the method itself (Fig.~\ref{fig:characterisation}) and to draw the expected ensemble trajectories in both panels of Fig.~\ref{fig:validation}, since a run of only 77 games is too noisy to draw them (the Equal treatment has just six games); the empirical curves in that figure use all 77 games, and the scores of Table~\ref{tab:val} come from the \emph{matched run} only. 

\begin{algorithm}[tb]
\caption{Empirical heuristic-ensemble CRD simulation (single heuristic or ensemble).}
\label{alg:sim}
\begin{algorithmic}[1]
\For{each treatment $\omega$ and its number of games}
  \For{each game}
    \State initialize common fund, endowments, contribution history
    \For{$k = 1$ to $R$}
      \For{each player $i$}
        \If{$k=1$} \State draw $c^{\,\omega}_{1,i}$ from the empirical first-round distribution
        \ElsIf{configuration $=$ ensemble}
          \State evaluate all heuristics; assemble decision vector $\bm{c}^{\,\omega}_{k,i}$
          \State select $c^{\,\omega}_{k,i}$ by plurality vote --see Eq.~(\ref{eq:plurality})
        \Else
          \State set the conditioning value for the chosen heuristic; draw $c^{\,\omega}_{k,i}$ from $P_h(c\mid\text{state})$
        \EndIf
        \State update spent endowment and contribution history
      \EndFor
      \State update cumulative common fund
    \EndFor
    \State record per-player outcomes (treatment, endowment, contributions, payoff)
  \EndFor
\EndFor
\end{algorithmic}
\end{algorithm}

\subsection*{Ensemble decision rule (plurality arbitration)}\label{sec:ensemble}
In the ensemble configuration, at each round $k$ every player $i$ evaluates all heuristics, each yielding a hypothetical contribution $c^{\,\omega,\text{h}}_{k,i}\in\{0,2,4\}$~MU. These are assembled into a decision vector
\begin{equation}
\bm{c}^{\,\omega}_{k,i}=\bigl(c^{\,\omega,\mathrm{self}}_{k,i},\,c^{\,\omega,\mathrm{peer}}_{k,i},\,c^{\,\omega,\mathrm{fund}}_{k,i},\,c^{\,\omega,\mathrm{budget}}_{k,i}\bigr),
\end{equation}
whose components may coincide (agreement) or differ (conflict); with four heuristics and three actions the vector spans $3^4=81$ combinations. The executed contribution is the most frequent value of the decision vector (its mode); in case of a tie between modes, a single value is drawn uniformly at random from the tied options:
\begin{equation}
c^{\,\omega}_{k,i}=
\begin{cases}
\operatorname{mode}\bigl(\bm{c}^{\,\omega}_{k,i}\bigr), & \text{if a unique mode exists},\\[3pt]
\operatorname{Uniform}\bigl(\operatorname{modes}(\bm{c}^{\,\omega}_{k,i})\bigr), & \text{otherwise.}
\end{cases}
\label{eq:plurality}
\end{equation}
This is the standard hard-voting (plurality) aggregation rule \cite{Black1948,Mulligan2018}. It is a non-parametric scheme: behavioural dominance emerges from the decision process itself, with no pre-defined strategy proportions in the population or additional assumptions. Each component corresponds to a distinct projection of the game state, own past action, peers' past actions, collective progress, and own resource state, which in the worked example are exactly the four feedback channels the experimental interface displayed after each round.
Figure~\ref{fig:vote} works through one round of this rule, including a two-two tie resolved by the random draw.

We use a plurality vote for two reasons. It adds no parameters, and it keeps the rule easy to read: the executed action is just the one that most heuristics point to. Other choices exist. Fitting a single model jointly on all four state variables is the most direct alternative, and weighted or softmax votes are others; we did not test these here. It is also worth saying where the method sits relative to familiar ones. Unlike learning models such as experience-weighted attraction \cite{camererho1999ewa} or reinforcement learning \cite{erevroth1998reinforcement}, the heuristics do not update inside a game; each is a fixed conditional distribution read off the data. And unlike the Strategy Frequency Estimation Method \cite{Dalbo2011,fudenberg2012slow}, which infers how common each strategy is in a population, we fix no proportions in advance and let the vote settle each round. 

\section*{Method validation}\label{sec:validation}

The method is validated following the \emph{pattern-oriented modelling} strategy \cite{grimm2005pattern}, under which a model earns structural credibility by reproducing multiple independent patterns at different scales rather than by fitting a single curve. As we have stressed above, in the \emph{matched run} we simulate the same number of games and treatment configurations as the target dataset and compare synthetic and empirical macro-indicators chosen to span the individual/collective and static/dynamic dimensions, so that no indicator trivially encodes the logic of any heuristic.  Every comparison is restricted to goal-reaching games, in the data and in the simulations, so the final payoff $e^{\,\omega}_{R,i}$ of a player is the endowment kept. Here, $\omega$ indexes the treatment, $k$ the round and $i$ the player, $e^{\,\omega}_{0,i}$, $c^{\,\omega}_{k,i}$ and $\phi^{\,\omega}_{k}$ are the quantities of the game-setting subsection, and the subscripts/superscripts ``$\mathrm{emp}$'' and ``$\mathrm{syn}$'' mark empirical and synthetic values where applicable. Similarity is quantified with a metric matched to each indicator's mathematical object:
\begin{itemize}
  \item \textbf{Final payoff distribution} (individual, static): the \textbf{Jensen--Shannon divergence} \cite{lin1991divergence}, is symmetric, finite even on disjoint support, and bounded. For each endowment level $\epsilon$, let $p_\epsilon$ and $q_\epsilon$ be the empirical and synthetic histograms of $e^{\,\omega}_{R,i}$ (bins of 2~MU, a constant $10^{-8}$ added to every bin before renormalising) and $m_\epsilon=\tfrac12(p_\epsilon+q_\epsilon)$; then
  \begin{equation}
  \mathrm{JS}_\epsilon=\tfrac12\sum_{b}p_\epsilon(b)\ln\frac{p_\epsilon(b)}{m_\epsilon(b)}+\tfrac12\sum_{b}q_\epsilon(b)\ln\frac{q_\epsilon(b)}{m_\epsilon(b)}.
  \label{eq:js}
  \end{equation}
  It is paired with the \textbf{Wasserstein-1 (earth-mover's) distance} \cite{rubner2000earth}, a true metric on distributions that is robust to binning. We compute it on the normalised payoff $u=e^{\,\omega}_{R,i}/(e^{\,\omega}_{0,i}/2)$, so it is dimensionless and $u=1$ means that the player kept half of her endowment:
  \begin{equation}
  W_{1,\epsilon}=\int_0^1\left|F^{-1}_{p,\epsilon}(t)-F^{-1}_{q,\epsilon}(t)\right|\mathrm{d}t,
  \label{eq:w1}
  \end{equation}
  with $F^{-1}$ the quantile function of $u$.
  \item \textbf{Final Lorenz curve / inequality} (collective, static): the absolute difference in \textbf{Gini coefficients} and the \textbf{area between the synthetic and empirical Lorenz curves}. Lorenz curves are monotone and indexed by population fraction rather than time. For each treatment, $\bar L^{\,\omega}$ is the mean Lorenz curve of the final payoffs over its games, and $G^{\omega}=1-2\int_0^1\bar L^{\,\omega}(u)\,\mathrm{d}u$ (trapezoidal rule):
  \begin{equation}
  \Delta G^{\omega}=\left|G^{\omega}_{\mathrm{syn}}-G^{\omega}_{\mathrm{emp}}\right|,
  \qquad
  A^{\omega}=\int_0^1\left|\bar L^{\,\omega}_{\mathrm{syn}}(u)-\bar L^{\,\omega}_{\mathrm{emp}}(u)\right|\mathrm{d}u.
  \label{eq:lorenz}
  \end{equation}
  \item \textbf{Average contribution per round} and \textbf{common-fund trajectory} (individual/collective, dynamic): the \textbf{normalised mean-squared error} (nMSE, pointwise level fidelity) together with \textbf{dynamic time warping} (DTW)~\cite{sakoe1978dynamic} (tolerant to phase and ramp-speed differences). Both temporal trajectories use the same metric pair. The average contribution of each endowment level (50 points over the five levels, in the order 30, 60, 48, 24 and 40~MU) and the mean common fund of each treatment (30 points over Equal, Unequal-H and Unequal-L) are each concatenated into a single vector $y$ of length $n$, and one score is computed on it. DTW is the cost $D(n,n)$ of the optimal warping path between $y^{\mathrm{emp}}$ and $y^{\mathrm{syn}}$,
  \begin{equation}
  D(a,b)=\left|y^{\mathrm{emp}}_{a}-y^{\mathrm{syn}}_{b}\right|+\min\left\{D(a{-}1,b),\,D(a,b{-}1),\,D(a{-}1,b{-}1)\right\},
  \label{eq:dtw}
  \end{equation}
  with $D(0,0)=0$ and $D(a,0)=D(0,b)=\infty$ for $a,b>0$, computed with the FastDTW approximation of the \texttt{fastdtw} Python package (radius 1). Being a sum along the path, it grows with $n$ and is therefore compared across models within a row of Table~\ref{tab:val}, not across rows.
\end{itemize}

Throughout, nMSE denotes the mean-squared error between the synthetic and empirical trajectories normalised by the variance of the empirical trajectory, 
\begin{equation}
\mathrm{nMSE}=\frac{\tfrac1n\sum_{j=1}^{n}\left(y^{\mathrm{emp}}_j-y^{\mathrm{syn}}_j\right)^2}{\mathrm{Var}\!\left(y^{\mathrm{emp}}\right)},
\label{eq:nmse}
\end{equation}
so a value of 1 corresponds to an error the size of the data's own variability. Moreover, unless stated otherwise, each value reported below is an unweighted mean across endowment levels for the payoff metrics and across the three treatments for the inequality metrics, while the two temporal indicators are single scores on the concatenated vectors described above. This matches the aggregation used in the source study, so the two sets of numbers can be read side by side.

\begin{figure}[htbp]
\centering
\includegraphics[width=\textwidth]{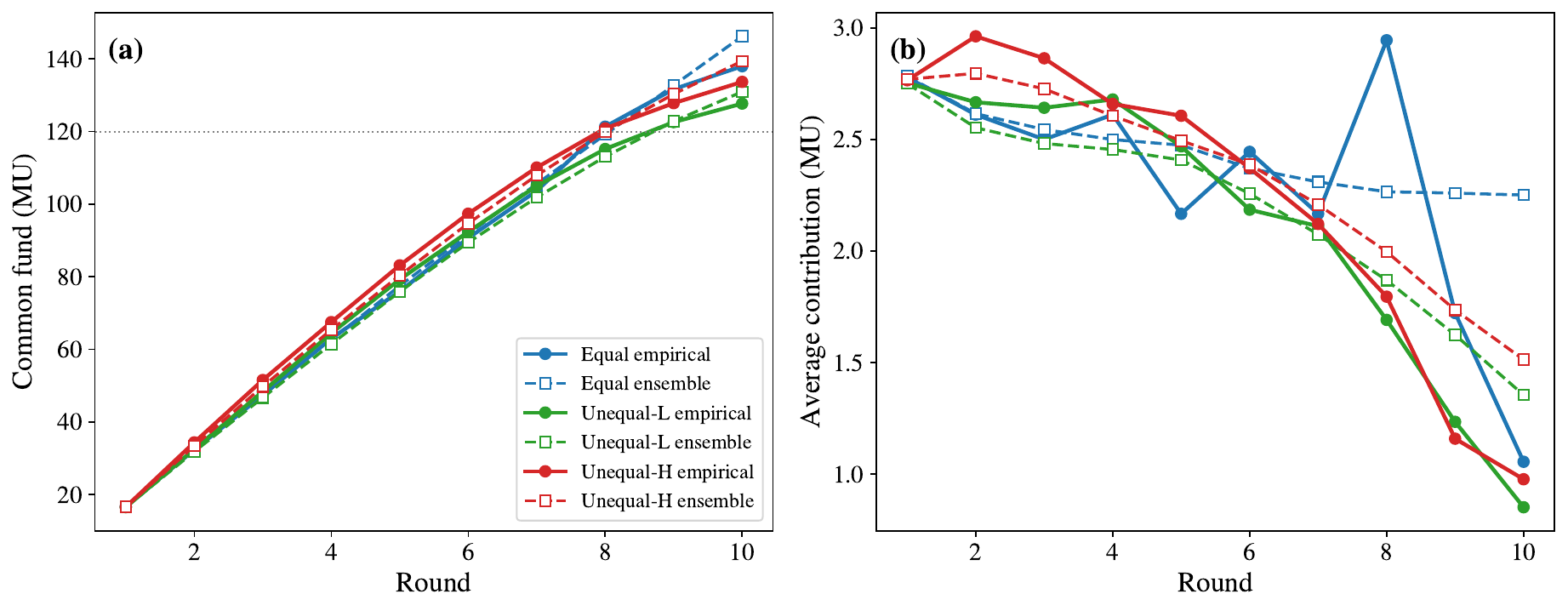}
\caption{Method validation on the worked example, \emph{matched run}, dynamic indicators. (a) Common-fund trajectory and (b) average contribution per round, per treatment, for the empirical data (solid) and the ensemble (dashed). The dotted line in (a) marks the target (120~MU). The static indicators (payoff distribution and inequality) are quantified in Table~\ref{tab:val}. Closeness indicates how well the method reproduces each empirical pattern.}
\label{fig:validation}
\end{figure}

To make these numbers interpretable, each metric is reported not only for the ensemble but also for (i) a uniform-random null in which agents draw actions uniformly from $\{0,2,4\}$~MU allowed by their remaining endowment and (ii) the four single-heuristic models, so that the ensemble's fidelity is read relative to chance and to its own components. Because one 77-game random draw is noisy, the null is the mean of independent replicates of the matched run. Table~\ref{tab:val} and the overlays in Fig.~\ref{fig:validation} report the validation against the worked-example dataset. One number in Table~\ref{tab:val} needs a caveat. The common fund is reproduced almost exactly, at nMSE $0.010$ for the ensemble, but that closeness is partly built in. The heuristics are seeded on empirical contribution probabilities, so the total contributed over ten rounds is already near the data before any arbitration happens, and matching the integrated fund is therefore close to guaranteed. The demanding tests are the round-resolved and distributional patterns. There the ensemble does well but not perfectly: the average-contribution nMSE is $0.283$, and Fig.~\ref{fig:validation}(b) shows the clearest gap, the late-round fall in contributions, which the ensemble smooths over rather than reproduces. The ensemble is closer to the data than the mean of the random null on all eight indicators, by 1.6 to 2.2 standard deviations of the null on the payoff and inequality indicators and by 4.2 to 7.5 on the dynamic ones. This is not true of every single heuristic: on average $H_{\mathrm{peer}}$ and $H_{\mathrm{budget}}$ are no closer than chance on the two inequality indicators, so neither carries, on its own, information about how final wealth is dispersed. The scalar Gini difference is a coarse number that can be small by coincidence, which is why we also report the Lorenz-area metric, which compares the whole curve; the two rank the five models in the same order.

\begin{figure}[htbp]
\centering
\includegraphics[width=\textwidth]{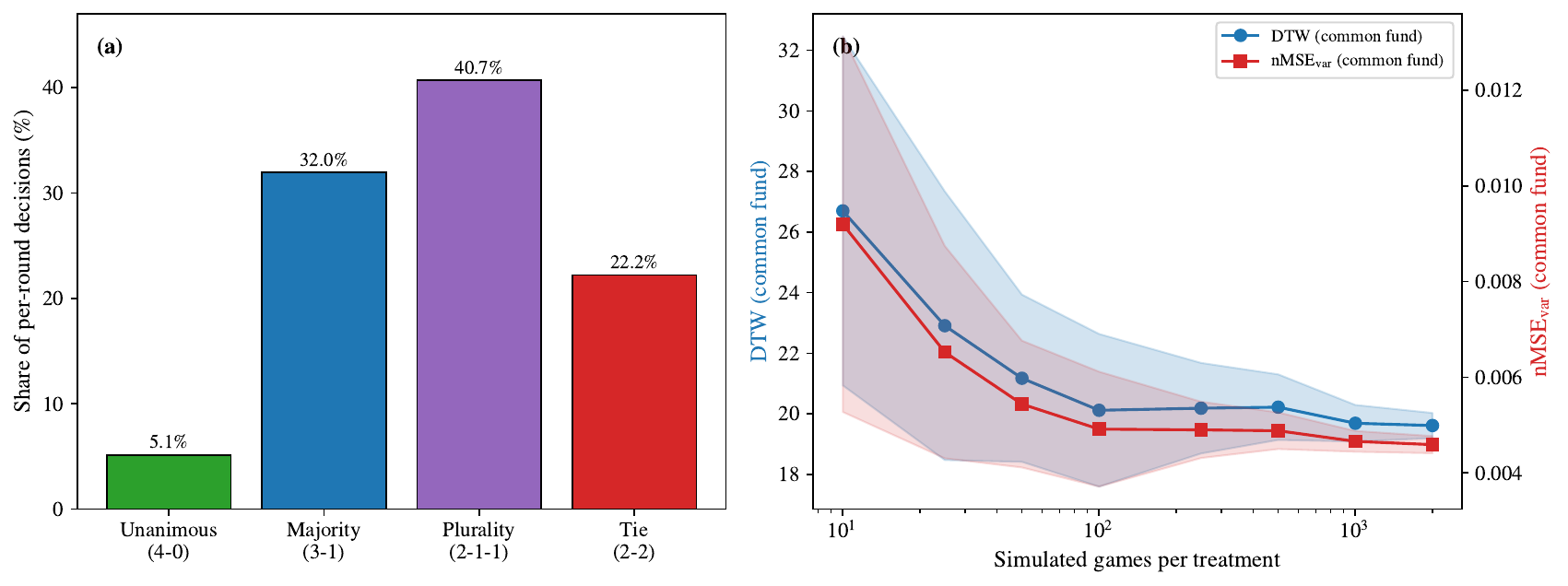}
\caption{Characterisation of the worked-example ensemble, \emph{extensive run}. (a) How the four heuristics agree. (b) The two common-fund similarity scores as a function of simulated games per treatment: Dynamic Time Warping (DTW, left axis) and variance-normalised nMSE (right axis). Shaded bands are $\pm1$ standard deviation across Monte-Carlo repetitions and narrow as more games are drawn.}
\label{fig:characterisation}
\end{figure}

Two further properties of the worked-example instance are given in Fig.~\ref{fig:characterisation}, generated from the \emph{extensive run}: how often the plurality vote is decisive for this set of four heuristics, and how the similarity scores settle as the number of simulated games grows. As shown in Fig.~\ref{fig:characterisation}(a), strong majorities, where three or four heuristics coincide, are only about $37\%$ of decisions ($5.1\%$ unanimous and $32.0\%$ three-to-one). A further $40.7\%$ are three-way splits decided by a single vote, and the remaining $22.2\%$ are two-two ties settled by the random draw. The vote is therefore decisive most of the time, but a large share of decisions is only weakly determined by the heuristics. Figure~\ref{fig:characterisation}(b) shows the two common-fund scores, DTW and nMSE, computed here per treatment and averaged over the three treatments, and therefore on a different scale from the pooled values of Table~\ref{tab:val}. Both settle once a few hundred games per treatment are simulated. At the sizes of the matched run, 25 to 50 games per treatment in the figure, they still fluctuate by about 13 to 34\% (one standard deviation) around their converged value, so differences between models in Table~\ref{tab:val} that are smaller than that should not be over-read. Together with the validation scorecard of Table~\ref{tab:val}, these results characterise how the method behaves and how much sampling uncertainty its scores carry.

\begin{table}[htbp]
\centering
\caption{Validation scorecard of the \emph{matched run}: similarity between synthetic and empirical macro-indicators for the four single heuristics, the ensemble, and a uniform-random baseline. Lower divergence/error/distance means closer reproduction; the best value \emph{among the models} (Random excluded) is in bold. nMSE is variance-normalised mean-squared error. No single heuristic reproduces every pattern: individual heuristics perform particularly well for specific indicators, whereas the ensemble remains competitive across all dimensions. 
}
\label{tab:val}
\resizebox{\textwidth}{!}{%
\begin{tabular}{l c c c c c c}
\toprule
Macro-indicator (metric) & $H_{\mathrm{self}}$ & $H_{\mathrm{peer}}$ & $H_{\mathrm{fund}}$ & $H_{\mathrm{budget}}$ & Ensemble & Random \\
\midrule
Payoff distribution (JS)            & \textbf{0.069} & 0.106 & 0.087 & 0.115 & 0.098 & 0.123 \\
Payoff distribution (Wasserstein-1) & \textbf{0.084} & 0.124 & 0.120 & 0.133 & 0.123 & 0.157 \\
Inequality ($|\Delta$Gini$|$)       & 0.056 & 0.071 & 0.066 & 0.074 & \textbf{0.047} & 0.067 \\
Inequality (Lorenz area)            & 0.028 & 0.036 & 0.033 & 0.038 & \textbf{0.023} & 0.035 \\
Average contribution (nMSE)         & 0.725 & 0.813 & \textbf{0.185} & 0.552 & 0.283 & 1.065 \\
Average contribution (DTW)          & 19.263 & 19.136 & \textbf{9.357} & 17.317 & 11.158 & 23.897 \\
Common fund (nMSE)                  & 0.080 & 0.051 & 0.013 & 0.053 & \textbf{0.010} & 0.128 \\
Common fund (DTW)                   & 138.033 & 119.148 & \textbf{75.297} & 121.077 & 85.228 & 144.174 \\

\bottomrule
\end{tabular}
}
\vspace{3pt}

\begin{minipage}{\textwidth}
\footnotesize
\textit{Note:} Every column is restricted to goal-reaching games. Random is the mean of 195 independent replicates of a 77-game uniform-random run, in which actions are drawn from the subset of $\{0,2,4\}$~MU allowed by the remaining endowment (5 of 200 replicates had no goal-reaching game in some treatment or endowment level and were discarded); its standard deviation across replicates is 7 to 18\% of the mean, and it is chance \emph{conditioned on success}, a conservative floor rather than unconditional chance. The two dynamic rows are pooled scores over concatenated curves, so DTW is comparable across models within a row but not across rows. $|\Delta$Gini$|$ is a coarse scalar and can be small by coincidence; it is read alongside the Lorenz-area metric, which compares the full curves.
\end{minipage}
\end{table}

\section*{Limitations}
The conditional distributions are estimated from a single dataset; applying the framework to a new dataset requires re-estimating them, and the worked-example validation does not by itself establish transfer to other populations or game structures. The four heuristics are one instantiation, chosen for the observable state of the worked-example game; other datasets may require a different set of heuristics, which the framework admits by construction. The discretization of conditioning variables introduces a bin-width choice that affects the estimated probabilities. Plurality voting is one of several possible arbitration rules; weighted, softmax, or priority-ordered are possible alternatives. The round-by-round fit is the weaker side of the ensemble. Because each heuristic reads only the previous round, the ensemble cannot generate a coordinated end-game drop in contributions, which is what the data show in the final rounds. It tracks the build-up of the fund well and misses the late fall. The large-scale simulations are generative and are not validated against empirical data beyond the matched-size comparison reported above.

\section*{Ethics statements}
The data used to instantiate and validate this method were collected in accordance with the Declaration of Helsinki, within the citizen science programme of the OpenSystems group of the Universitat de Barcelona, and approved by the Universitat de Barcelona Ethics Committee (IRB00003099) approved this mobility experiment. All participants read and signed the informed consent form and parental/legal guardian consent, if appropriate. No privacy issues were observed to be in conflict with the public release of the underlying processed data. Data were used in anonymised form \cite{Vicens2018xairedata}.

\section*{CRediT author statement}

{\bf M. Sadurn\'i:} Conceptualization, Methodology, Software, Validation, Formal analysis, Investigation, Data Curation, Writing - Original Draft, Writing - Review \& Editing, Visualization. {\bf M.F. D\'iaz:} Methodology, Software, Validation, Formal analysis, Data Curation, Writing - Original Draft, Writing - Review \& Editing, Visualization. {\bf J. Perell\'o:} Conceptualization, Methodology, Investigation, Resources, Data Curation, Writing - Review \& Editing, Supervision, Project administration, Funding acquisition. {\bf M. Montero} Conceptualization, Methodology, Investigation, Resources, Writing - Review \& Editing, Supervision, Project administration, Funding acquisition.

\section*{Declaration of competing interest}
The authors declare that they have no known competing financial interests or personal relationships that could have appeared to influence the work reported in this paper.

\section*{Acknowledgements}
The work was partially supported by MICIU/AEI/10.13039/501100011033, grant number PID2019-106811GB-C33 [MS, JP and MM]; by MICIU/AEI/ 10.13039/501100011033 and by ``ESF Investing in your future'', grant number PRE2020-093266 [MS]; by MICIU/AEI/10.13039/501100011033 and by “ERDF/EU”, grant numbers PID2022-140757NB-I00 and PID2025-171705NB-I00 [MS, MFD, JP and MM]. We also acknowledge the support of Generalitat de Catalunya through Complexity Lab Barcelona, grant number 2021SGR00856 [MS, MFD, JP and MM].

\section*{Declaration of Generative AI and AI-assisted technologies}

During the preparation of this work the authors used Claude (by Anthropic) in order to improve the language, grammar, syntax, and clarity of the manuscript. All scientific content, methodological decisions, analyses, and conclusions were conceived and written primarily by the authors. After using this tool, the authors reviewed and edited the content as needed and take full responsibility for the content of the publication. 
\section*{Data and code availability}
The source code implementing the method is available at \href{https://github.com/mf-diaz/Project-xAire-EHE}{Project-xAire-EHE} (GitHub repository). Starting only from the deposited data, it regenerates Tables~\ref{tab:firstcon} and~\ref{tab:val} and the data-driven Figs.~\ref{fig:heatmap}, \ref{fig:validation} and~\ref{fig:characterisation}, including the synthetic runs of the four heuristics and of the ensemble, and it checks the outcome against stored reference values. The repository also holds the figures of the article. Without the 30{\,}000-game extensive run the pipeline takes about two minutes; with it, about two hours on one core. The empirical data are not duplicated in the repository. The worked-example dataset is available in the companion data article \cite{Vicens2018xairedata} and in the CORA repository \cite{DATA3672_2026}.

\bibliographystyle{elsarticle-num}
\bibliography{references}
\end{document}